\documentstyle[emlines]{article}

\begin{document}
\large
\section{Introduction}

A dissipation free inertial range of scale is assumed to exist  in a
theory  of developed turbulence  of incompressible fluid.
Under the scenario of isotropic homogeneous turbulence the infrared (IR)
pumping of energy into this range is executed by the large  vortices of
scale $L$; the small scale
$ l_{0} $  defines the bottom of the range, where the dissipation
of energy because of viscosity becomes significant.

The attempts to generalize  the approach advanced in \cite{1}-\cite{2}
 on a  case of  weak compressibility of a fluid was established in
\cite{3}-\cite{4}. It was shown, that since some values of Mach number
 $Ma = v_c/c$ ($v_c$ is the velocity of turbulent pulsation,
 $\quad Ó$ means the velocity of sound)  a new  parameter $c^{-2}$ occurs
in the inertial range. In the case of large Mach numbers it makes
 impossible to construct a theory of developed turbulence with the help
of dimensionless parameters.

Under such circumstances the most important problem of the theory is to
 describe the scaling functions. The general method of construction of
 decomposition of IR  asymptotics of Green functions on  $Ma$ number
(zero order of decomposition is an incompressible fluid) was advanced
in \cite{5}. To study the strength of the method we offer the adaptation
of it  for a model of stochastic  magnetic hydrodynamics (MHD) in the
frameworks of present paper.

The original results have been obtained by us about the renormalization and
critical dimensions of the composite operators (the local averages) of
 fields in magnetic  hydrodynamics are  significant because of it is
 available for an experimental measurement.

\section{Model of stochastic magnetic hydrodynamics of  compres\-sible
fluid}

As initial micromodel a system of equations of stochastic magnetic
hydrodynamics, completed by the equation of continuity
and condition of weak compressibility of a fluid is chosen.
\begin{equation}
\partial_t{\bf \varphi}=-(\varphi\partial){\bf \varphi}+\frac 1{\rho}
 \left(\eta \Delta {\bf\varphi} + \eta'{\bf\partial}(\partial\varphi)-
{\bf\partial}p-\frac 12 {\bf\partial}\theta ^2
+ (\theta\partial) { \bf \theta } \right ) + { \bf f^\varphi },
\end{equation}
\begin{equation}
\partial_t { \bf \theta } = - ( \varphi\partial ) { \bf \theta } + \nu_m
\Delta { \bf \theta } + ( \theta\partial ) { \bf \varphi } - { \bf \theta } (
\partial \varphi ) + { \bf f^\theta },
\end{equation}
\begin{equation}
 \partial _t \rho + { \bf \partial}(\rho {\bf\varphi})=0,\quad
 p=c^2\rho.
\end{equation}
The Navier-Stokes equation (1) for the velocity of fluid
${\bf \varphi}({\bf x},t)$ contains the terms are responsible for
interaction with the pseudovector ${\bf \theta}({\bf x},t)=
{\bf B}/\sqrt { 4\pi\rho_0 }$. We assume here $\rho_0$ as an equilibrium
value of density of fluid (further we accept $\rho_0 =1$);
$p({\bf x},t)$ is a field of  pressure, $\eta$ is a factor of shift viscosity,
 $\eta'= ( \eta / 3 + \zeta ),$  where $\zeta$ is a factor of volumetric
viscosity, $c$- velocity of sound. The magnetic viscosity coefficient
 $\nu_m=c^2 / 4\pi\sigma =\lambda\nu$  (here  $\sigma$ means the conductivity of fluid,
$c$ is the  speed of light ) is useful to be connected with a factor of
kinematic viscosity with the help of inverse Prandtl number $\lambda$.

For the external random force ${\bf f}^\varphi$ referred
 to unit of weight and a rotor of the random current ${\bf f}^\theta$ it is
supposed Gaussian distribution with the correlators  $D_ {ij}^{\phi\phi}$
\cite{5}:
\begin{equation}
 D^{\varphi\varphi} _ { ij } = { g_1 } _0 \nu^3_0 P_ { ij } d_ { vv } +
 c^{-4}{g_4}_0Q_{ij}d_{uu}.
 \quad
 D^ { \theta\theta } _ { ij } = { g_2 } _0 \nu^3_0 P_ { ij } d_ {
 \theta\theta }, \quad D^ { v\theta } _ { ij } = { g_3 } _0 \nu^3_0
\varepsilon_ { ijm } k_m d_ { v\theta },
\end{equation}
$$
 d_ { vv } =d_ { uu }
= k^ { 4-d-2\varepsilon },\quad d_ { \theta\theta } = k^ { 4-d-2 a\varepsilon
},\quad d_ { v\theta } = k^ { 3-d- ( 1 + a \varepsilon)}.
$$
($D_ { ij } $ do not depend of frequency), $d$ - dimension of space,
(generally speaking the antisymmetric pseudotensor $\varepsilon_{ism}$
 is determined only at $d=3$).

The factors ${g_i}_0$ in (4)  play a role of charges; the positive constant
$a$ is an arbitrary parameter of theory.
 $\varepsilon$ is served to construct the decomposition of correlation
functions. Really $\varepsilon_p = 2$; it simulates the energy pumping from
the large-scale movements of a fluid.

In the framework of approach it is useful to assume
$g_3=\xi\sqrt{g_1g_2}.$ A new dimensionless parameter $\xi$ replacing
  $g _3$ is  not a charge but an arbitrary ($|\xi|\leq 1$) parameter of
 the theory.

A generating functional of the correlation functions of the stochastic
problem (1-4) is defined by the formula
 $$
 G ( A_\phi ) =\int { \cal D } \Phi \det M \exp ( S ( \Phi ) +
\Phi A_\Phi ),\quad \Phi=\ { \varphi,\varphi',\theta,\theta', p, p'\ },
 $$
in  which  the action functional  $S(\Phi)$  is a kind of
\begin{equation}
\begin{array}{c}
S ( \Phi ) = 1 / 2 \varphi'_iD^ { \varphi \varphi } _ { ij } \varphi'_j + 1 /
 2 \theta'_iD^ { \theta\theta } _ { ij } \theta'_j + \varphi'_iD^ {
 \varphi\theta } _ { ij } \theta'_j + \varphi'_i\Bigl [ -\partial_t \varphi_i
- ( \varphi\partial ) \varphi_i + \\
 + 1 / ( 1 + p / c^2 ) \Bigl ( \eta\Delta { \bf \varphi } _i
+ \eta'\partial_i ( \partial \varphi ) -\partial_i p - 1 / 2\partial _i
 \theta^2 + ( \theta\partial ) \theta_i \Bigr ) \Bigr ] + \theta'_i\Bigl [
 -\partial_t \theta_i + \\
 + \lambda\nu \Delta \theta_i - ( \varphi\partial )
 \theta_ i + ( \theta\partial ) \varphi_i- \theta_i ( \partial \varphi )
 \Bigr ] + p'\Bigr [ 1 / { c^2 } \partial _t p + ( \partial u
 + 1 / { c^2 } \partial ( p\varphi ) \Bigr ].
\end{array}
\end{equation}

Necessary integration on ${\bf x}$, $t$ and summation on indices are meant.
 $A_\Phi$ are the sources of appropriate fields; $\det M$
is an unessential factor under conditions of the theory  \cite{3}.

A perturbation theory  for  $G(A_\Phi)$ is received under the decomposition
of $\exp S$ on the terms of (5) containing the products of three and more
fields. $1/(1+p/c^2)$  should be presented as the power series of $p / c^2.$
The correlation functions of an arbitrary number of $ \rho\equiv p/c^{2} $
 are divergent in the theory, so the infinite number of counterterms
 would be required for renormalization of (5). Therefore the application of
renormalization group approach (RG) to (5) is impossible.

At $c^{-2}=0$ integration on the field $p'$ of the generating functional
yields $\delta (\partial\varphi)$. It corresponds to the case of
incompressible fluids $\rho \to \rho_0.$ By the way a field of velocity
becomes  $v_i =P_{ij}\varphi_j, \quad P_ {ij}=\delta_ {ij}
-k_ik_j / k^2$, and the longitudinal component $u_i= Q_ { ij }
\varphi_j, \quad Q_ { ij } = \delta_ { ij } -P_ { ij } $ is excluded from
stochastic equations (1-2). The appropriate theory contains the transversal
fields only: $v,\quad v'$ and $\theta,\quad\theta'.$ For  elimination
of divergences in such theory a final number of counterterms is required only.

The theory of incompressible  conductive fluid was investigated in \cite{6}
in the first order on $\varepsilon$ with the help of recursive renormalization
group, and then under the  field theory approach in \cite{7}. The scaling
behaviour of correlation functions in such model was considered in \cite{8},
and the critical dimensions of the composite operators of junior canonical
dimensions were calculated in \cite{9}.

Hereinafter it will be convenient pursuant to estimation of $u\sim c^{-2}$
\cite{5} to make following replacement of variables in (5):
$$ u\to u / c^2,\quad p'\to c^2 p'. $$

The action (5) is invariant to a Galileian transformations of fields:
$$
{\bf\varphi}({\bf x}, t)\to{\bf\varphi}({\bf x}+{\bf a} t, t)-{\bf a},
\quad \Phi({\bf x}, t)\to\Phi({\bf x}+{\bf a} t, t),
$$
where $\Phi$ are the other fields, except of ${\bf\varphi},$ which
participates in (5) whether under a derivative or in the covariant derivative
${\cal D} _t = \partial_t + v\partial.$

\section{IR perturbation theory in a kinetic mode of compres\-sible fluid}

 The action functional (5) develops from the  incompressible one
$S_{ic}(v,v',\theta,\theta')$ and the terms  containing the
longitudinal and scalar fields  $u, u', p, p',$. For $S_ { ic } $ the
IR-behaviour  was studied by RG. The RG transformation  has two  fixed points
$$
{\beta_g} _\alpha=0,\quad\omega_ {\alpha\delta}\equiv\partial{\beta_g}_\alpha
/\partial g_\delta >0,\quad{\beta_g}_\alpha\equiv D_M
g_\alpha,\quad D_M\equiv M\partial_M,
$$ ($M$ is a renormalization mass parameter) determining two different
critical regimes of turbulent behaviour \cite{8}. So called ''kinetic'' regime
realizes if a trajectory of invariant charges $\bar g_\alpha$ falls into a
vicinity of a point
\begin{equation}
{g'_1}_{*} =\frac{{g_1}_{*}}{B\lambda_{*}}=\frac{\varepsilon(1+
\lambda_{*})}{ 15 },\quad {g'_2}_{*} =\frac {{g_2 }_{*}}{B\lambda_{*}^2}
 =0, \quad \lambda_{*} = \frac {\sqrt{43/3}-1}{2},
\end{equation}
in the  charge space. $B=d(d+2){(4\pi)}^{d/2}\Gamma(d/2)$.
The point is IR steady while $a < 1.16.$

A magnetic regime is been steady at $a\geq 0.25$ to be correspond to a point
\begin{equation}
{g'_1}_{*}=\lambda_{*}=0,\quad{g'_2}_{*}= a\varepsilon.
\end{equation}

Because of trivial values of invariant charges under (6) conditions it is
useful to apply the new variables \cite{8}:
\begin{equation}
\theta\rightarrow\sqrt{g_2\nu^3}M^{a\varepsilon}\theta,\quad \theta'
\rightarrow\theta'/\sqrt{g_2\nu ^3}M^{a\varepsilon},\quad
v\rightarrow\sqrt{g_1\nu^3}M^\varepsilon v,\quad v'\rightarrow v'/\sqrt
{g_1\nu^3}M^\varepsilon
\end{equation}
to avoid triviality of the correlation functions in the kinetic regime. The
position of the fixed point (6) doesn't change under (8).

After integration of $G(A_\Phi)$ on the transversal fields $v, v',\theta,
\theta'$ one can construct a new perturbation theory  on $c^{-1}$
as a kind of decomposition of $\exp(S-S_{ic})$ on the terms containing the
longitudinal and scalar fields.  The appropriate diagrammatic techniques
consists of lines of fields $u, u', p, p'$, and the composite operators
 of $v, v',\theta,\theta'$. The propagators of
longitudinal and scalar fields join own vertices or connect to the composite
operators of transversal fields.

For summation of IR peculiarities in subgraphs are build of transversal
fields one can use the results \cite{5}, \cite{9} where critical dimensions
of the fields and some composite operators are indicated. Dimensions of a
 number  of composite operators containing the magnetic fields
 $\theta$ and $\theta',$ have been unknown and are resulted here for the
first time.

 One can coordinate the diagrammatic techniques with decomposition on $c^{-1}$
with the help of a change of variables in (5):
\begin{equation}
\tilde q=p + g_2\nu^3\theta^2 / 2-F_p,\quad \tilde u_i=u_i+{F_1}_i-{F_2}_iq,
\end{equation}
where $F_p=\nu^3\Delta^ { -1 } \partial_i\partial_k ( g_1
v_iv_k-g_2\theta_i\theta_k ),\quad { F_1 } _i = \Delta^ { -1 } \partial_i {
\cal D } _tF _p,\quad { F_2 } _i= \Delta^ { -1 } \partial _i { \cal D } _t$,
and $\quad\quad \Delta^{-1}$ is the Green function  of Laplasian.

 A problem of renormalization of operator sequences containing a number of
$F_p$-type operators had been  discussed already in \cite{5}. The analysis
of them was facilitated by  the Galileian invariance of theory, since
the noninvariant local operators  constructed with participation of  field
$v$ do not admix to  a combination $g_1\nu^3\partial_i\partial_kv_iv_k$
accumulated in the perturbation theory \cite{5}.

In our case the set of  operators mixed due to renormalization
 appears  to be essentially wider owing to Galileian invariancy of
 $\theta$-field. Renormalization of such sequences demands  the knowledge
 of scaling dimensions of the local composite operators each, which we
have not. Therefore, discussing critical behaviour in MHD we are compelled
to be limited to approach of one insertion of the operators of transversal
fields.

As a result of  transformation (8) one can get:
\begin{equation}
\begin{array}{c}
S(\Phi)=1/2v'_id^{vv}_{ij}v'_j+1/2\theta'_id^{\theta\theta}_{ij}
\theta'_j+\xi v'_id^{v\theta}_{ij}\theta'_j+1/2 u'_iD^{uu}_{ij}u'_j
+v'_i\Bigl[-\partial_t v_i -\\
-\nu^{3/2}{g_1}^{1/2}(v\partial)v_i-
1/{c^2}\left((v\partial) \tilde u_i + (\tilde u\partial)v_i\right ) +
1/{(1 + \tilde q/c^2)}\Bigl(\nu\Delta v_i+\\
+g_2{g_1 }^{-1/2}\nu^3 (\theta\partial ) \theta_i +  1 / { c^4 }
 \nu'\nu^ { -3 / 2 } g_1^ { -1 / 2 } \tilde q\Delta\tilde u_i\Bigr ) \Bigr ]
+u'_i[-1/{c^2}\partial _t\tilde u _1/{c^2}g_1^{1/2}\nu^{3/2}\cdot \\
 \cdot\Bigl((v\partial)\tilde u_i+(\tilde u\partial)v_i\Bigr)-
 1/{c^4}(\tilde u\partial ) \tilde u _i-\partial _i p
+ 1 / ( 1 + \tilde q / c^2 ) ( - 1 / { c^2 } \nu^ { 3 / 2 } \nu g_1^ { 1 / 2 }
\tilde q\Delta v _i + \\ + 1 / { c^2 } \nu'\Delta \tilde u _i - 1 / 2\cdot 1 /
{ c^2 } \partial _i { \tilde q } ^2
- 1 / { c^2 } \nu^3 g_2\tilde q ( 1 / 2 \partial _i \theta ^2
- ( \theta\partial ) \theta _i )) ] + \\
+\theta'_i\Bigl[
 -\partial_t \theta_i + \lambda\nu \Delta \theta_i - g_1\nu^ { 3 / 2 } \left
(( v\partial ) \theta_ i + ( \theta\partial ) v_i\right ) + 1 / { c^2 } \left
 (( \theta\partial ) \tilde u _i- \theta_i ( \partial \tilde u - ( \tilde
u\partial ) \theta _i \right ) \Bigr ] -\\ -p'\Bigl [ ( \partial u + 1 / {
c^2 } \partial_i ( \tilde q\tilde u_i ) \Bigr ],\quad\nu'=(\eta+\eta')/\rho_0.
\end{array}
\end{equation}
Square-law part of the functional defines the propagators of  perturbation
theory:
\begin{equation}
\begin{array}{c}
G^{vv'}={(G^{v'v})}^{+}=L_1^{-1},\quad
G^ { \theta\theta' } = { ( G^ { \theta'\theta } ) } ^ { + } =L_2^ { -1 },\\
G^{vv}=G^{vv'}d_{vv}G^{v'v},\quad
G^ { \theta\theta } =G^ { \theta\theta' } d_ { \theta\theta } G^ {
\theta'\theta },\\ G^ { \phi'\phi' } =0,\quad G^ { uu' } = { ( G^ { u'u } ) }
^ { + } =\omega L_3^ { -1 },\quad G^ { p'p } = { ( G^ { pp' } ) } ^ { + } =- (
i\omega + \nu'k^2 ) L_3^ { -1 }, \\ G^ { uu } =G^ { uu' } D_ { uu } G^ { u'u
},\quad G^ { pp } =G^ { pu' } D_ { uu } G^ { u'p },\quad G^ { up } =1 / c^2 G^
{ uu' } D_ { uu } G^ { u'p }, \\ G^ { p'u } =i { \bf k } / k^2,\quad G^ { u'p
 } =i { \bf k } { L_3^ { * }} ^ { -1 }, \\ L_1=-i\omega + \nu_0 k^2,\quad
L_2=-i\omega + \lambda_0\nu_0 k^2, \quad
\end{array}
\end{equation}
The propagators of transversal fields are multiple to $P_ { ij }$;
 other ones have the factor of $Q_ { ij }.$

The scaling dimensions of operators of transversal fields are listed in
table 1. In the second column of (tab. 1)  the composite operators of
transversal fields are submitted. The longitudinal and scalar fields
joined with them  are specified in column 4. The degree  of $c^{-2}$
at an appropriate vertex is indicated in column 5.  The third column of
the table gives a scaling dimension of the appropriate composite operator
at the physical value of $d=3,\quad \varepsilon=2$. In the table we have
taken $\omega_{1,2}=\min(\omega_1,\omega_2)$ and $\omega_{1,3}=
\min(\omega_1,\omega_3).$ The IR-correction indices
($\omega_\alpha$) are the eigenvalues of a matrix $\omega_{\alpha\delta}$.
They are known to be $\omega_\alpha >0$ to provide the stability of a
critical regime.

\section{Scaling dimensions of the composite operators of trans\-versal
 fields}

One should note  that the contributions of composite operators constructed
with participation of fields $(g_2\nu^3)^{1/2}\theta$ to the asymptotics of
Green functions  in the perturbation theory are proportional to degrees
of $\bar g _2.$ Within the framework of RG the estimation of value of
$\bar g _2$ in the vicinity of  ${g_2}_{*}$ is proven:
$$ \bar g_2=a_2 s^{\omega_2}$$
($s$ is a dimensionless scaling parameter; $a_2$ is an arbitrary constant).
So the index $\omega_2$ have to be accounted at the decision of critical
dimension of the operators of type mentioned.

The renormalization of composite operators of transversal fields proceeds
in the framework of incompressible theory with the renormalized functional
of action
$$
\begin{array}{c}
S _R= \frac12 v' d_{vv}v' + \frac12 \theta' d_{\theta\theta} \theta' +
 v'  d_{v \theta} \theta' +\\
+ v' \bigl [ -\partial _t  v + Z _1\nu \Delta v
-{g_1}^{1/2}\nu^{3/2}M^\varepsilon ( v \partial) v + Z
 _3g_2{g_1}^{-1/2}\nu^{3/2}M^{(2a-1)\varepsilon}
 (\theta \partial)\theta \bigr]+ \\
+ \theta' \bigl [ -\partial _t  \theta + Z _2\lambda\nu \Delta \theta
-{g_1}^{1/2}\nu^{3/2}M^\varepsilon ( v \partial) \theta
+ {g_1}^{1/2}\nu^{3/2}M^\varepsilon(\theta
\partial)v \bigr],
\end{array}
$$
where the one-loop order of renormalization constants were calculated
in \cite{7}:
$$
\begin{array}{c}
Z _1 =
1-\frac{g_1d(d-1)}{4B\varepsilon}-
\frac{g_2(d^2+d-4)}{4Ba\lambda^2\varepsilon},\quad
Z _3 =
1+\frac {g_1}{B\lambda\varepsilon}-\frac {g_2}{Ba\lambda^2\varepsilon},\,
\nonumber\\
Z _2 =
1-\frac{g_1(d
+2)(d-1)}{2B\lambda(\lambda+1)\varepsilon}-
\frac{g_2(d+2)(d-3)}{2Ba\lambda^2(\lambda+1)\varepsilon}.
\end{array}
$$

\subsection{Not local composite operators}

It is known, that the local operators (i.e. respected to a same point) can
admix to not local one as the result of renormalization. The operators we are
interested in the Galileian invariant operators would be admixed only.
 All the not local composite operators in the  case considered contain
the single  insertion of $F_p$-type. They have two external derivative;
the admixing local operators should have the same property. As far as
$d_\partial=1$, one can show that all such local operators are unessential
in comparison with the not local ones \cite{5}. Therefore it is available to
limit the consideration of matrix of renormalization constants by the
 set of elements which are responsible for mixing of not local operators.
The appropriate scaling dimensions are the sums of scaling dimensions of
local fragments, derivatives, and $ \Delta^{-1} $-operators. As a matter
of fact it is sufficient to study the scaling dimensions of the local
fragments.

\subsection{Local composite operators}

The dimension of a tensor $\phi_i\phi_k$ was defined in \cite{9}.
In case of small $a$ the operator $\partial_i\partial_k v_{i}v_{k}$
is essential, its dimension is known precisely ($4 / 3$) at real value of
$\varepsilon$. Depending on the value  of parameter  $a$ the contribution of
 operator $g_2\partial_i\partial_k \theta_i\theta_k$  can occur essential
too ($5\frac 15-4a + \omega_2$ ). We know only one-loop value of index
$\omega_2\simeq 4(a-1.16)$ \cite{7}, so it is possible to predict shift of
dimension $\Delta F_p$ at $a\geq 1.1,$ i.e. near the boundary  of area of
stability  of kinetic regime.

The dimension of local fragments located in $N^0 3,4$ in tab.1
can be determined, having considered result of action
of the operations $\partial_{g_\delta },$ $g_\delta=g_1, g_2,\lambda,$
on a RG equation  for renormalized generating functional of spanned Green
functions $W^R(A_\Phi)$:
$$ D_ { RG } W^R ( A_\Phi ) =0,\quad D_ { RG } =D_M + \beta _{g_\alpha}
\cdot\partial_ { g_\alpha } -\gamma_1\cdot D_\nu,
\quad \gamma_\alpha= D_M \ln Z_{\alpha}.$$

The linear combinations of operators $C_\alpha= {\omega_{\delta\alpha}}^{-1}
[\partial_{g_\delta}, D_{RG}]W$ have in fixed points (6,7) anomalous
dimensions $\omega_{\alpha}$ (square brackets here designate a commuta\-tor).

In a fixed point is fair
$ D_ { RG } C_\alpha=\omega _ { \alpha\delta } C_\delta,$ as far as
$$
\begin{array}{c}
D_ { RG } C_\alpha= \omega_ { \delta\alpha } ^ { -1 } D_ { RG } ( \omega_ {
\delta\alpha } \cdot \partial_ { g_\alpha } - \partial _ { g_\delta }
\gamma_1\cdot D_\nu ) W = \Bigl ( D_ { RG } \cdot\partial_ { g_\alpha } - \\ -
\omega_ { \delta\alpha } ^ { -1 } D_ { RG } \partial _ { g_\delta }
\gamma_1\cdot D_\nu
\Bigr)W
=\Bigl (( \omega_ { \delta\alpha } \partial _ { g_\delta } - \partial _ {
g_\delta } \gamma_1\cdot D_\nu ) - \omega_ { \delta\alpha } ^ { -1 } \cdot
\beta _ { g_\alpha } \partial_ { g_\alpha } \partial _ { g_\delta } \gamma_1
D_\nu \Bigr ) W=\\ = \omega _ { \alpha\delta } C_\delta.
\end{array}
$$
In the last equality we allowed, that in the fixed point $\beta_{g_\alpha}
=0$. Eigenvalues $\omega_{\alpha}$ matrixes $\omega_{\alpha\delta}$
 are the anomalous dimensions of combinations $C_\alpha.$

In a vicinity  of  kinetic point (7) the matrix has a kind of:
\begin{displaymath}
\omega_{\delta\alpha}=\left(
\begin{array}{ccc}
3 g_1\partial_ { g_1 } \gamma_1 & 3 g_1\partial_ { g_2 } \gamma_1& 0 \\
 0& -2a\varepsilon + 3\gamma_1-\gamma_3& 0 \\
 0 &\lambda \partial_{g_2}\gamma_1& -\lambda \partial_\lambda\gamma_2\\
\end{array}
\right).
\end{displaymath}
The eigenvalues $\omega _\alpha$ of  matrix  coincide with the diagonal
 elements.

Taking into account, that in kinetic  point it is fair:
$\partial_ { g_\delta } \beta_ { g_1 } =3 g_1\partial_ { g_\delta}
 \gamma_1,$
one receives  the implicit expressions for $C_\alpha$ :
$$ C_\alpha= [ \partial_ { g_\alpha } -\frac1 { 3 g_1 } \delta_ { \alpha 1 }
\cdot D_\nu ] W. $$
One can make out them obviously:
$$
\begin{array}{l}
C_1=(Z_1/{2\varepsilon g_1} \cdot (\gamma_1-2\varepsilon/3))v'\Delta
v + (Z_2/{2\varepsilon g_1} \cdot (\gamma_2-2\varepsilon/3)) \lambda
\theta'\Delta\theta; \\
C_2=v'\Delta v + \chi v' (\theta\partial) \theta,
\quad {\chi} = {(\partial_{g_2} Z_1)} ^ {-1} Z_3 {g_1}^{1/2}
\nu^{1/2} M^{(2a-1) \varepsilon}
 \\ C_3= \lambda(1+\lambda\partial_{\lambda})Z_{2}\theta'\Delta\theta.
\end{array}
$$
(Here we have taken advantage by that in the first order of $\varepsilon$
$\quad \gamma_i=2\varepsilon g_1\partial_ { g_1 } Z_i$.)

The operators $v'\Delta v,\quad \theta'\Delta \theta$ and $v' (
\theta\partial ) \theta$ are contained in the action (11);
dimensions of  $C_\alpha$ are:
\begin{equation}
\Delta_ { C_1 } =3\frac 23 + \omega_1, \quad\Delta_ { C_2 } =3\frac 23 +
\omega_2, \quad\Delta_ { C_3 } =3\frac 23 + \omega_3
\end{equation}
Essentiality of the contributions of particular combinations depends on a
parity between the correction indexes $\omega_\alpha$ and is defined
by the least of them.

Renormalization of the operators ($N^0 1$, tab.1)
proceeds in a class of tensors at a zero external momentum and frequency:
$$
{\bf{\cal G}}=\left\{\partial(v'v),\quad\partial(\theta'\theta),
\quad v'_i\partial_j v_i,\quad\theta'_i\partial_j \theta_i\right\}.
$$
The dimensions of two first operators of family ${\bf{\cal G}}$ is
$\Delta=4$, as far as these operators don't renormalize and don't admix
to any others, as a structure of interaction in the theory of type
\cite{1}, \cite{7} provides removal one derivative on each external line
$v'$ and $\theta'$ - types from each 1-irreducible diagram, that effectively
lowers an index of divergence of the diagram.

 Two last operators in ${\bf{\cal G}}$ are the vectors
  not equal to zero at a zero external momentum.
The operator ${\bf{\cal G}}_1=v'_i\partial_j v_i$ is finite.
One can represent it in a kind of a difference of two finite operators:
$\partial_j ( v'_iv_i ) -v_i\partial_j v'_i.$ (The finiteness of
$v_i\partial_jv'_i$ in the Galileian invariant theories is proven in
 \cite{5}. This operator is Galileian invariant, therefore it can
 not admix to ${\bf{\cal G}}_1$). By virtue of finiteness  of
${\bf{\cal G}}_1$,  operator ${\bf {\cal G}}
_2 =\theta'_i\partial_j\theta_i$ does  not admix to it.

As a whole, the matrix of renormalization constants has a kind of:
\begin{displaymath}
{\bf Z} =\left(\begin{array}{cc}
1&Z_{12}\\
0&Z_{22}
\end{array}
\right).
\end{displaymath}
In one-loop order approach the unknown elements of matrix $Z_{ik}$
 are calculated under the diagrams listed on firure 1. It gives:
\begin{equation}
Z _ { 12 } =
\frac { \lambda ( d + 2 ( d-1 ) } { ( \lambda + 1 } ( \frac { g'_1 } {
\varepsilon } - \frac { g' _2 } { a\varepsilon } ), \quad Z _ { 22 } =
1-\frac { ( d-1 ) ( d + 2 } { 2 ( \lambda + 1 } ( \frac { g'_1 } {
 \varepsilon } + \frac { g'  _2 } { a\varepsilon } ).
\end{equation}
With the known matrix $Z_{ik}$ a matrix of critical dimensions $\Delta_{ik}
=(d_F^k)_{ik}+\Delta_\omega (d_F^\omega)_{ik}+\gamma_{ik}$ is calculated.
Critical dimensions of frequency in the incompressible theory \cite{8}
are $\Delta_\omega=2 / 3$ in kinetic  regime and $\Delta_\omega=2$ for
 magnetic one. The concrete critical dimensions there are eigenvalues of
 matrix $\Delta_ {ik}.$ They are corresponded to the linear combinations
 $U_ { ik } F_k^R$ making  $\Delta_{ik}$ diagonal.
Critical dimensions belong to the operator ${\bf{\cal G}}_2,\quad
\Delta_ {{ \bf { \cal G }} _2 } =5\frac 13$, as well
as combination $ { \bf { \cal G }} _2 + \mu { \bf { \cal G }} _1,\quad \Delta_
{{ \bf { \cal G }} _2 + \mu { \bf { \cal G }} _1 } =4,\quad \mu =\gamma_ { 12
} \gamma_ { 22 } ^ { -1 }.$

It is important to note that the scaling behaviour in the theory is
determined by not renormilized composite operators. The relation between
the renormalized combination and the not renormilized ones is followed:
$$
  F_{\delta}^{R}=U_{\delta\alpha}Z_{\alpha\beta}^{-1}F_{\beta}.
$$
Thus we have
$$
 {\bf {\cal G}}_{1}^{R}+\mu{\bf{\cal G}}_{2}^{R}={\bf {\cal G}}_{1} +
(\mu-Z_{12})Z_{22}^{-1} {\bf{\cal G}}_{2}, \quad {\bf{\cal G}}_{2}^{R}=
Z_{22}^{-1} {\bf{\cal G}}_{2}.
$$
The most essential contribution connected with $ { \bf { \cal G }} _2$ is
given by a linear combination of the operators $ { \bf { \cal G }} _1$ and
$ {\bf { \cal G }} _2$; it has the dimension $\Delta=4$ ( $N^0 1$, tab. 1).

The dimension of some composite operators  can be solved without
calculating of the appropriate diagrams with the help of Schvinger equations.
One can consider for example equality of a kind:
$$0=\int{\cal D}\Phi\frac{\delta}
{\delta v'_i(x)}\{g_2\theta^2(x)e^{S_R+A\Phi}\}.$$
In designations $ <F>_{A} \equiv
\frac { \int D\Phi F ( x,\Phi ) e^ { S_ { R } ( \Phi ) + A\Phi }} { \int D\Phi
e^ { S_ { R } ( \Phi ) + A\Phi } } $ it will be recorded as
\begin{equation}
< G_2\theta^2 (x) V_ { i } (x) >_ { A } =-g_2 A_ { v '}(x) < \theta^2 (x) >_
{ A },
\end{equation}
where
$$
\begin{array}{c}
V_ { i } \equiv \frac { \delta S_ { R } ( \Phi ) } { \delta v ' _ { i } (x) }
 = ( d_ { vv } v ' ) _ { i } + ( d_ { v\theta } \theta' ) _ { i } - { \cal D }
 _ { t } v _ { i } + \nu Z_1\Delta v _ { i } + \\ + Z_3 g_2 g_1^ { -1 / 2 }
 \nu^ { 3 / 2 } M^ { ( 2 a-1 ) \varepsilon } ( \theta\partial ) \theta_i-
 g_2\nu^3\partial _ { i } \theta^2 / 2 + \partial_iF_p.
\end{array}
 $$
The dimension of object in a left-hand part of equality (14) is defined
as follows:
\begin{equation}
\Delta _ { FV_i } =\Delta _F -\Delta _ { v'} -\Delta_t-
\Delta _x,\quad  F= { g_2\theta^2 }.
\end{equation}
 It should be noticed, that the contribution of random forces in $V_{i}$
 $$\theta^2 {(d_{v\theta}\theta')}_i + \theta^2 { ( d_ { vv } v' )
 } _i \equiv
\theta^2 (x) \left ( \int dx d^ { vv } _ { ij } ( x-y ) v'_j (y) + \int dy d^
{ v\theta } _ { ij } ( x-y ) \theta'_j (y) \right ) $$
is finite and doesn't admix to any other considered operators
because of presence of closed cycles of retarded lines in 1-irreducible
diagrams. Hence, it can be cancelled.

In a kinetic regime  at the real value of $\varepsilon$ the critical
dimensions are $\Delta _ { x } =-d,\quad \Delta_t=-2 / 3,\quad\Delta_ { v' }
=3\frac 13 $ \cite{7}. Then, with the accounting of (15) one can
receive for a combination $ { \cal J } _1=
g_2\theta^2V_i$:
 $$ { \Delta _ { \cal J }} _1=2\frac 13-4 a + \omega_2. $$

An equation
\begin{equation}
0=\int{\cal D}\Phi\frac{\delta}{\delta v'_i(x)}
\{g_2v'\theta^2(x)e^{S_R+A\Phi}\},
\end{equation}
is useful for a determination of dimension of a combination
$ { \cal J } _2=v'\theta^2 ( - { \cal D } _tv + \nu \Delta v -
g_2\nu^3\partial\theta^2/2+\partial F_p+g_2g_1^{-1/2}\nu^{3/2}(\theta\partial
 )\theta)$

With the preceding notation the equation has a kind of:
$$ < g_2\theta^2 (x) V_ { i } (x)>_{A}+(d-1)g_2\delta^{(d + 1}(x-y)|_{x=y}
 \theta^2 (y) = -g_2 A_ { v ' } (x) < \theta^2 (x) >_ { A }.
$$
 Formally infinite term in a left-hand part of equality containing
$\delta^{(d+1)}(0)$  will be cancelled with the similar
ultraviolet divergent  contribution on the average $\\<g_2\theta^2(x)V_i(x)>_A$;
at calculations in the frame of network of minimal subtraction (MS)  they can
 be simultaneously ignored. Then the  formula (15) permits to receive
the required  dimension, if a dimension of the operator $X=g_2v'_i\theta^2$
is known.

To study the dimension it is necessary to analyse renormalization of
family of the operators:
$$ { \cal X } =\ { v'_iv^2,\quad g_2 v'_i\theta^2, \quad \theta'_i \theta v\
}. $$
The operators of ${\cal X}$ don't  mix with the operators of
 $\phi'\partial \phi$ - type,  as far as the 1-irreducible diagrams
 do not contain divergences.
The last operator of family $ { \cal X } $ is not mixed with the other
operators for the same reason. Furthermore the Galileian invariancy of the
considered theory causes finiteness of the operators $v'_iv^2$,
$ \theta'_i \theta v$ ($\Delta=4$) and the absence of admixing of $v'_iv^2$
 to  $v'_i\theta^2.$

It is obviously that the  critical dimension  inherent in kinetic mode to
combina\-tion of the operators
${\bf X}={\cal X}_1^R+\mu_3 {\cal
X}_2^R={\cal X}_1+(\mu_3-Z_{12})Z_{22}^{-1}{\cal X}_2,$
 $\mu_3=\gamma_{12}\gamma_{22}^{-1}$ is $\Delta_{\bf X}=2\frac 23.$
Independently the operator ${\cal X}_2^R=Z_{22}^{-1}{\cal X}_2$ has
a critical dimension too.  Appropriate  constants $Z_{12}$ and $Z_{22}$
 (determining the anomalous $\gamma_{ik}$) in one-loop approach
can be  calculated on diagrams located on figure 2.
$$
Z_{12}=-\frac{\lambda(d+2)(d-1)}{2(\lambda + 1)}(\frac{g'_1}{\varepsilon}
-\frac{g'_2}{a\varepsilon}),$$
$$
Z_{22}Z_\theta ^2 = 1 + \frac{(d-1)(d+2)}{2(\lambda + 1)}(\frac{g'_1}
{\varepsilon}-\frac{g'_2}{a\varepsilon}). $$
So for critical dimension of  ${\cal X} _2$ at the real values of
 $\varepsilon$ and $d$ one can receive:
$$\Delta_{{\cal X}_2} =6\frac 23- 4 a. $$

The most essential contribution, connected with the operator
 $g_2v'_i\theta^2$  is $\Delta= 2\frac 23 +\omega_2$
 (at $a<1$) (stipulated by a combination ${\bf X}$) and
 $\Delta =6\frac 23-4a + \omega_2$ (in the case of  $a>1$) (when
the contribution of ${\cal X}_2^R$ becomes essential).

In view of this result from the formula (15) is received scaling
 dimension for the combination ${\cal J}_2$:
$$
\Delta_{g_2v'_i\theta^2V_i}=
\Bigl\{\begin{array}{c}
 5\frac 23-4 a + \omega_2,\quad a>1 \\
 1\frac 23 + \omega_2,\quad a<1.
        \end{array}
\Bigr.
$$

 \section{Results and Discussion concerned with the kinetic re\-gime}

A dimensionless variable $k^2c^2/\omega^2$ in (11) defines the scaling
dimension of $c$ as $\Delta_c=-1/3,$ as well as in the theory of ordinary
compressible fluid \cite{5}. Then with the propagators (11) it is possible
 to determine of scaling  dimension of longitudinal and scalar fields:
\begin{equation}
\Delta_p=-2 / 3,\quad \Delta _ { p' } =3\frac 23,\quad \Delta_u=-1,\quad
\Delta_ { u' } =3\frac 13.
\end{equation}
In the framework of Landau ideas to get an ''effective''
functional of action the developed turbulent spectra in the kinetic mode
are described, one should reject all  the unessential operators.
Problem about whether the operator is essential or not we  decide proceeding
 from its scaling dimension calculated at real values of $\varepsilon$ and
 $d.$ Results of comparisons of dimensions of the operators are shown in
tab. 2.

There are the operators having the same order on $c^{-2}$ and connected in
functional (10) with identical sets of longitudinal and scalar fields are
compared. In the first column the numbers of lines of table 1 are specified,
where the essential operators of smaller dimension are located; in
the second column the unessential operators are displayed.
The conditions at which comparison of the operators is executed are
indicated in the third column of the table.

The analysis of data of table 2 shows that depending on size of parameter
 $a$ whether the composite operators containing the not local insertion
of $F_p$-type   or the  operators constructed with participation of local
insertion $g_2\theta^2$ (9) are essential.
The boundary value of $a$ at which shift of a critical mode occurs is
$a=2/3 + \omega_2/4 =5/6-1/8{\bf\gamma}\simeq 0.91.$
 As was already specified down to $a\geq 1.1$ the leading contribution
 in $F_p$ is connected with the combination ${\cal R}_2$ Thus, the
 contribution  of $\partial_i\theta_k\partial_k\theta_i$ in kinetic regime
 of the model are  correction always.

To get the ''effective'' functional of action depending on value of
parameter $a$ we should exclude the various unessential operators. Resulting
functional has a form of:
\begin{equation}
\begin{array}{c}
S(\Phi) = 1 / 2 v'_id^ { vv } _ { ij } v'_j +
 1 / 2 \theta'_id^ { \theta\theta } _ { ij } \theta'_j + \xi v'_id^ { v\theta
} _ { ij } \theta'_j + 1 / 2 u'_iD^ { uu } _ { ij } u'_j + v'_i [ - { \cal D }
 _t v_i-\\
- 1 / { c^2 } \left (( v\partial ) \tilde u_i + ( \tilde u\partial ) v_i
\right)]
+ \theta'_i [ - { \cal D } _t \theta_i + \lambda\nu \Delta \theta_i -
g_1\nu^{3/2}(\theta\partial)v_i+
 1 / { c^2 } (( \theta\partial ) \tilde u _i + \\ + \nu \Delta v_i-\theta_i (
 \partial \tilde u ) ]
+ u'_i [ - 1 / { c^2 } { \cal D } _t\tilde u _
 1 / { c^2 } g_1^ { 1 / 2 } \nu^ { 3 / 2 } ( v\partial ) \tilde u_i + \nu' /
c^2\Delta u_i + ( \tilde u\partial ) v_i-\\ -1 / c^2 \nu'\partial_i\partial_t
q + 1 / c^2 \Delta^ { -1 } \partial_i\partial^2_tq -1 / { c^4 } ( \tilde
u\partial ) \tilde u _i-\partial _i p + \\
+ 1 / 2 ( 1 + \tilde q / c^2 ) \cdot 1 / { c^2 } \partial _i { \tilde q } ^2 ]
-p'\Bigl [ ( \partial u + 1 / { c^2 } \partial_i ( \tilde q\tilde u_i ) \Bigr
],
\end{array}
\end{equation}
where we have used the  notations followed:
\begin{equation}
Q\to p + g_2\nu^3\Theta ( a-2 /3-1/ 4\omega_2 ) \theta^2 / 2,\quad\tilde
q=q-F_p, \quad \tilde u_i = u_i + { F_1 } _i- { F_2 } _iq,
\end{equation}
and $F_p=g_1\nu^3\Delta^ { -1 } \Theta ( 2 / 3 + 1 / 4\omega_2-a )
\partial_iv_k\partial_k v_i, \quad { F_1 } _i = \Delta^ { -1 } \partial_i {
\cal D } _tF _p, \quad { F_2 } _i= \Delta^ { -1 } \partial _i { \cal D } _t$,
and $\quad\quad \Theta (x) $ is theta function.

The amendment of the first order on $c^{-2}$ to spectra of developed
turbulence in kinetic regime in the area  of $a>2/3 + \omega_2/4$
is given by the operator ${\cal D}_tF_2\theta^2$
($N^0 30$, tab. 3): $\sim c^{-2}k^{-2( 2a-\omega_2/2-1)}.$

We have defined  the scaling dimension of $c$ as $\Delta_c=-1/3.$
 Now it is visible that in accordance with results of the theory in the
framework of approach of one insertion of the composite
operators of transversal fields our estimation is fair at
 $a<2/3+\omega_2/4.$ In case of strong singularity of magnetic pumping of
energy the scaling behaviour of  $c$ becomes more essential
 $\sim k^{1 +\omega_2/2-2a}.$
The dimension  of pressure field $p$ thus should be  evaluated on
an insertion of operator $g_2\theta^2 / 2,\quad \Delta _p=2 + 1 /
2\omega_2-4 a$. And we have:
\begin{equation}
\Delta _ { p' } =1 + 4 a-\omega_2,\quad \Delta_u=1\frac 23 - 4 a +
\omega_2,\quad \Delta_ { u' } =\frac 23 + 4 a-\omega_2.
\end{equation}
The new variable appearing thus in the inertial range
is proportional $Ma$ $k^{1-2a+\omega_2/2}$.
The amendments to spectra of developed turbulence due to compressibility of
a fluid are essential in  the scaling area of the spectrum of scales.

We shall notice that the  results yielded are received by us in approach
 of one insertion of operators  of transversal fields, so it
would be changed in the case of accounting of more than one insertion.

\section{ On the IR perturbation theory in magnetic regime }

 In the magnetic fixed point (7)  all the invariant charges approaches to
zero. For the formulation  the nontrivial perturbation theory it
needs to be redefined fields, parameters, and the variable of time:
\begin{equation}
\begin{array}{c}
\theta\rightarrow\sqrt { \lambda } \theta,\quad \theta'\rightarrow 1 / \sqrt {
\lambda } \theta',\quad v\rightarrow\lambda v, \quad v'\rightarrow v',\quad
t\rightarrow t / \lambda, \\
u\rightarrow\sqrt{\lambda}u,\quad
P\rightarrow\sqrt { \lambda } p, \quad u'\rightarrow \sqrt { \lambda } u',
\quad p'\rightarrow \sqrt { \lambda } p', \quad c^ { -2 } \rightarrow \sqrt {
\lambda } c^ { -2 }.
\end{array}
\end{equation}
We shall note, that the redefinition of transversal fields in (21) doesn't
change a position of the magnetic fixed point (7) in the theory  of
incompressible fluid, \cite{8}. The action functional agreed with
 decomposition on $c^{-2}$ with the preceding notation
 $\tilde q$, $\tilde u$ will be recorded as
{\normalsize
\begin{equation}
\begin{array}{l}
S ( \Phi ) = 1 / 2 Bg'_1\nu^3 v'_id^ { vv } _ { ij } v'_j + 1 / 2
 Bg'_2\nu^3\theta'_id^ { \theta\theta } _ { ij } \theta'_j + B\xi v'_id^ {
v\theta } _ { ij } \theta'_j + 1 / 2\lambda u'_iD^ { uu } _ { ij } u'_j +
 v'_i\Bigl [ -\lambda\partial_t v_i - \\
-\lambda(v\partial)v_i
- 1 / { c^2 } \lambda\left (( v\partial ) \tilde u_i + ( \tilde u\partial )
v_i\right ) + 1 / { ( 1 + \lambda\tilde q / c^2 ) } \Bigl ( \nu\Delta v_i + (
 \theta\partial ) \theta_i + 1 / { c^4 } \lambda\tilde q\Delta\tilde u_i\Bigr
 ) \Bigr ] + \\
+\theta'_i\Bigl[
 -\partial_t \theta_i + \lambda\nu \Delta \theta_i - \left (( v\partial )
\theta_ i + ( \theta\partial ) v_i\right ) + 1 / { c^2 } \left ((
 \theta\partial ) \tilde u _i- \theta_i ( \partial \tilde u - ( \tilde
u\partial ) \theta _i \right ) \Bigr ] + \\ + u'_i [ - 1 / { c^2 } \lambda^ {
3 / 2 } \partial _t\tilde u _
 1 / { c^2 } \lambda^ { 3 / 2 } \left (( v\partial ) \tilde u_i + ( \tilde
u\partial ) v_i\right ) - 1 / { c^4 } \lambda^ { 3 / 2 } ( \tilde u\partial )
 \tilde u _i-\partial _i p + \\
+ 1 / ( 1 + \lambda\tilde q / c^2 ) ( - 1 / { c^2 } \lambda \nu\tilde q\Delta
v _i + 1 / { c^2 } \lambda^ { 1 / 2 } \nu'\Delta \tilde u _i
- 1 / 2\cdot 1 / { c^2 } \lambda \partial _i { \tilde q } ^2- \\
- 1 / { c^2 } \lambda^ { 3 / 2 } \tilde q ( 1 / 2 \partial _i \theta ^2
- ( \theta\partial ) \theta _i )) ]
-p'\Bigl [ ( \partial u + 1 / { c^2 } \lambda\partial_i ( \tilde q\tilde u_i )
\Bigr ].
\end{array}
\end{equation}
}
As the operators of shift for $\tilde q$, $\tilde u$ in (22) are chosen
$F_p=\lambda^ { 1 / 2 } \Delta^ { -1 } \partial_i\partial_k \\ ( \lambda
v_iv_k-\theta_i\theta_k ),\quad { F_1 } _i = \lambda \Delta^ { -1 } \partial_i
{ \cal D } _tF _p,\quad { F_2 } _i=\lambda \Delta^ { -1 } \partial _i { \cal
D } _t$.

The square-law part of (22) defines the propagators
 functions  in a kind of:
{\normalsize
\begin{equation}
\begin{array}{c}
G^ { vv' } = { ( G^ { v'v } ) } ^ { + } =1 / \lambda_0 L_1^ { -1 },
\quad G^ {\theta\theta' } = { ( G^ { \theta'\theta } ) } ^ { + } =L_2^ { -1 },
\quad G^ {vv } =G^ { vv' } D_ { vv } G^ { v'v }, \\
 G^ { \theta\theta}=G^{\theta\theta'}D_{\theta\theta}G^{\theta'\theta },
\quad G^ { uu' } = { (
G^ { u'u } ) } ^ { + } =\omega L_3^ { -1 },\quad G^ { p'p } = { ( G^ { qp' }
) } ^ { + } =- ( i\omega + \nu'k^2 / \lambda ) L_3^ { -1 }, \\
G^{uu}=G^{uu'}D_{uu}G^{u'u},\quad
G^{pp}=G^{pu'}D_{uu}G^{u'p},\quad
G^ { up } =1 / c^2 G^ { uu' } D_ { uu } G^ { u'p }, \\ G^ { p'u } =i { \bf k }
/ ( \lambda k^2 ), \quad G^ { u'q } =i { \bf k } { L_3^ { * }} ^ { -1 } /
\lambda, \\ \quad L_1=-i\lambda_0\omega + \nu_0 k^2,\quad
L_2=-i\lambda_0\omega +
\lambda_0\nu_0k^2,
\quad L_3=-i\lambda\omega + \nu'k^2 + i k^2 c^2 / ( \lambda\omega )
\end{array}
\end{equation}
}
The propagators of transversal fields are multiple to $P_ { ij }$;
 other ones have the factor of $Q_ { ij }.$

The singularities  of functions (23) on $1/\lambda$
are displayed in the $\delta$-functions at $\lambda\to 0.$
However, the infringement of correctness of the perturbation theory
does not occur as far as all the in  (22),
 except $\theta'_i\left((\theta\partial)\tilde u _i- \theta_i
(\partial\tilde u) - ( \tilde u\partial)\theta _i\right),$ contain
in factors the positive degrees of
parameter $\lambda.$ Having seen  the diagrams before  passage to
the limit $\lambda\to 0$ it is easy  to be convinced
that all the singularities in them are eliminated by integration on a time
in the vertices and are reduced by factors at interactions.

The graphs constructed with participation of vertices  \\
$\theta'_i \left ((\theta\partial ) \tilde u _i-
\theta_i ( \partial \tilde u) - ( \tilde
u\partial ) \theta _i \right )$ contain  functions of propagators
independent from $\lambda$, they haven't the singularities  of  specified
 type.

 In the expansion series for multipliers $1/(1 +\lambda\tilde q/c^2)$
   in the functional (22) there are the vertices with
 a number of scalar fields $p$ as well as with insertions of the
 composite operators of transversal fields $F_p,$ $\lambda\theta^2.$
 $\tilde q $  participates in these decompositions in a combination with
parameter $\lambda.$ Thereof it is possible to assert that appropriate
singularities are absent in all the orders of the perturbation theory.

 As follows from \cite{8} in a magnetic regime in the theory of
 incompressible  fluid ${\gamma_\nu}_{*}=0.$ Therefore in a
 magnetic mode the invariant variable of frequency $\bar\nu k^2/\omega$
 defines the dimension of $\omega$ as $\Delta _\omega=2.$ The
 behaviour of  $\nu'$ in the scaling area  is not known;
assuming $\nu'$ as scaling dimensionless parameter, we  receive
 $\Delta _\omega=2$ also.  Variable $ck/\omega$ in this case permits to
 evaluate the dimension of $c$ as $\Delta_c=1.$ Then the amendments
 due to compressibility to  spectra of turbulence in  the scaling area
 appear unessential in magnetic regime.

To justify this assumption it is necessary to find out,
what  scaling dimension does the parameter $c^{-2}$ demonstrate in
the diagrammatic contributions of the perturbation theory.
It is reasonable to consider  the leading scaling
 asymptotics without the accounting of correction indexes
$ \omega_{\alpha}. $

In this case after fulfillment of functional integration of $G(A_\phi)$ on
the transversal fields the set of diagrams of considered perturbation theory
will be consisted of the graphs with
the lines of longitudinal and scalar fields adjoined to the
 composite operators of transversal fields located in the table 3.

Renormalization of  the  operators  contained in the table had been
  discussed by us.
The dimension ($N^0 1,$ tab. 3) is inherent to the operator
$\theta'\partial \theta$ in the combination
 $\theta'\partial \theta + \lambda\mu_3 v'\partial
 v,$ where $\lambda\mu_3=\mu$. As far as formally in the magnetic point
 $\lambda_ { * } =0,$ the contribution of the hydrodynamic operator is
 absent. (With the accounting of (13) in  magnetic point (7) the eigen
dimension of operator $\theta'\partial \theta$ is equal to $4 + 20 a,$ so
the appropriate contribution is unessential.)

Thus, the leading scaling asymptotics in the theory of developed
turbulence of a compressible conductive fluid in a magnetic regime is given
by the action functional as follows:
{\normalsize
\begin{equation}
\begin{array}{l}
S ( \Phi ) = 1 / 2 v'_iD^ { vv } _ { ij } v'_j + 1 / 2 \theta'_i D^ {
\theta\theta } _ { ij } \theta'_j + \xi v'_iD^ { v\theta } _ { ij } \theta'_j
+ 1 / 2 u'_iD^ { uu } _ { ij } u'_j
+ v'_i\left ( -\lambda\partial_t v_i + \nu\Delta v_i\right ) + \\ + u'_i ( -1
/ c^2\cdot \lambda^ { 3 / 2 } \partial_tu_i + \lambda^ { 1 / 2 } \nu' /
c^2\Delta u_i-\partial_i q - 1 / c^2\cdot \lambda\nu'\partial_i\partial_tq + 1
/ c^2\cdot \lambda^2\Delta^ { -1 } \partial_i\partial^2_tq ) -\\ -p' (
\partial u + \theta'_i ( -\partial_t\theta_i + \nu\Delta\theta_i- ((
v\partial ) \theta_
(\theta\partial)v_i)+
1 / c^2 (( \theta\partial ) \tilde u _i- \theta_i ( \partial \tilde u - (
\tilde u\partial ) \theta _i ))
\end{array}
\end{equation}
}
After renormalization of transversal fields in (24) for appropriate
generating functional of correlation functions it is possible to develop
the standard perturba\-tion theory on formally small parameter $c^{-2}$
 for Green functions that are the intrinsic tensors with distinguished
 arguments of  time. In the contribution of each order of the perturbation
theory the  finite number of  diagrams are participated. For
construction of decomposition for the functions, containing odd number of
fields $\theta$ as well as functions not integrated on time, it is necessary
to consider the theory with respect of the correction indexes.
As is shown in \cite{9}, the scaling representations for such functions
contain an additional small factor of ${\cal O}(s^{\omega_\alpha}).$

The main amendments to Green functions  in the theory of
$\simeq c^ { -4 } k^4$ that justifies the assumption stated above about
critical dimensions   $\Delta_c=1$ and  $\Delta_t=-2$, and
about  the amendments due to compressibility of fluid are unessential.
 Formally these amendments are presented in a kind of power series  of
parameter $Ma$ $k.$ In the contrast with the kinetic regime here at any
value of $Ma$ the  parameter connected with the compressibility does not
penetrate into the inertial range and does not essentially influence on
turbulent behaviour in magnetic hydrodynamics.

\section{ Acknowledgments}

The author is grateful  to M.Yu. Nalimov and L.Ts.Adjemian for useful
discussion.

\newpage
{\small
{\bf Table 1}
\begin{tabular}{|c||c||c||c||c|}
\hline
\multicolumn{5}{|c|}{\sc Composite operators of transversal fields}
  \\ \hline
              \hline
\multicolumn{1}{|c|}{\bf N.}&
\multicolumn{1}{|c|}{\bf Operator}&
\multicolumn{1}{|c|}{\bf Dimension}&
\multicolumn{1}{|c|}{\bf Fields}&
\multicolumn{1}{|c|}{\bf Factor}  \\
  \hline
\it 1
&  $v'\partial v, \quad \theta'\partial \theta$&4&$u$&$
                  1/{c^2}$  \\             \hline
\it  2
&$\Delta   F_1{F_p}^l$&$1-2l/3$ &
$u',\quad q^n$ &$1/{c^{2(l+n+1)}}$  \\
                                                \hline
\it 3
&$v'\Delta v{F_p}^l $
&$ 3{\frac 23}- 2l/3+\omega_{1,2}$ & $q^n$
&$ 1/{c^{2(l+n)}}$ \\                          \hline
\it 4
&$\theta'\Delta \theta{F_p}^l $
&$ 3{\frac 23}- 2l/3+\omega_{1,3}$  & $q^n$
&$ 1/{c^{2(l+n)}}$ \\                          \hline

\it 5
&$\Delta F_2{F_p}^l$&
$ 1\frac 23 - 2l/3$&$ u',\quad q^{n+1}$ &$1/{c^{2(l+n+1)}}$ \\
                                                \hline
\it 6
&$ v'\Delta   F_1{F_p}^l,\quad (l+n)>0$&
$4 \frac 13-2l/3$&$q^n$&$1/{c^{2(l+n+1)}}$\\
                                                \hline
\it 7
&$\partial{F_p}^{l+1}$&$1-2(l+1)/3$&$u',
\quad q^n $& $ 1/{c^{2(l+n)}}$\\
                                                \hline
\it 8
&$v'\Delta {F_p}^l,\quad (l+n)>0$& $5 \frac 13- 2l/3$&$u,
\quad q^n $&$ 1/{c^{2(l+n+1)}}$ \\   \hline
 \it 9
 &$\partial{F_v}_1$&$0$&$p',\quad q$&$ 1/{c^2}$ \\
 \hline
\it 10
&$v'\Delta F_2{F_p}^l,\quad (l+n)>0$& $5- 2l/3$&
$q^{n+1}$&$1/{c^{2(l+n+1)}}$ \\
     \hline
\it 11
&$\partial F_2$&$ 2/3$&$p',\quad q^2$& $1/{c^2}$ \\
\hline
\it 12
&${\cal D}_t F_1$&$- 1/3 $&$ u' $&$ 1/{c^2}$ \\
\hline
\it 13
&$\partial F_p $&$ 1/3$ &$u,\quad p'$&$1/{c^2}$ \\
\hline
 \it 14
 &$F_1 \partial v$&$- 1/3$&$ u'$& $ 1/{c^2}$ \\
\hline
 \it 15
 &$F_1 \partial  F_p$&$- 2/3 $& $p'$&$ 1/{c^2}$ \\
\hline
 \it 16
 &${\cal D}_t F_2 $&$1/3$ &$q,\quad  u'$& $1/{c^2}$ \\
 \hline
\it 17
&$F_2 \partial  F_p $&$0$&$q,\quad  p' $&$ 1/{c^2}$ \\
\hline
 \it 18
 &$F_2 \partial  v$&$ 1/3$&$q,\quad  u'$&$1/{c^2}$ \\ \hline
  \it 19
&$\partial F_1$&$0$&$u,\quad u'$&$ 1/{c^4}$ \\ \hline
\it 20
&$\partial v$&$2/3$&$u,\quad u'$&$ 1/{c^2}$ \\ \hline
 \it 21
&$\partial F_2$&$ 2/3$&$q,\quad u,\quad u'$&$ 1/{c^4}$ \\
\hline
\it 22
&$F_2 \partial F_1$&$-1/3$&$ u'$&$ 1/{c^4}$ \\
\hline
\it 23
&$\Delta {F_p}^l $&$2- 2l/3$&$u,\quad q^n,\quad
u'$&$ 1/{c^{2(n+l+1)}}$ \\
\hline
\it 24
& $F_1 \partial F_2$&$- 1/3$&$q,\quad u'$&$  1/{c^4}$  \\
\hline
 \it 25
 &$ v'\partial v F_2,\quad\theta'\partial \theta F_2$&$3\frac 23$
 &$q$&$ 1/{c^2}$  \\         \hline
 \it 26
 &$F_2 \partial F_2$&$ 1/3$&$u',\quad q^2 $&$ 1/{c^4}$ \\
\hline
 \it 27
 &$ v' \partial vF_1,\quad \theta'\partial\theta F_1$&$3$& - &
 $ 1/{c^2}$ \\
\hline
\it 28 &$\Delta v {F_p}^l,\quad (l+n)>0$&$1\frac 23 - 2l/3$&$
 u',\quad q^n $&$ 1/{c^{2(l+n)}}$ \\
 \hline
\it 29
&$v'(\theta\partial)\theta F_p $
&$ 3+\omega_{1,2}+\omega_2$ & $q^n$
&$ 1/{c^{2(n+1)}}$ \\                          \hline
\it 30
&${\cal D}_t F_2 \theta^2$&$2\frac 13-4a+\omega_2$&
$u'$&$1/{c^2}$ \\
\hline
\it 31
&$\theta^2 {\cal D}_tv$&$2\frac 13-4a+\omega_2$&$u',\quad q^n$&$1/c^{2(n+1)}$ \\
\hline
\it 32
&$\theta^2 \Delta v $&$2\frac 13-4a+\omega_2$&$u',\quad q^n$&$1/c^{2(n+1)}$ \\
\hline
\it 33
&$\theta^2 (\theta\partial)\theta$&$2\frac 13-4a+\omega_2$&$u',\quad q^n$
&$1/c^{2(n+1)}$ \\
\hline
\it 34
&$\theta^2\Delta$&$4-4a+\omega_2$&$u',\quad u,\quad q^n$&$1/c^{2(n+1)}$ \\
\hline
\it 35
&$v'\theta^2 {\cal D}_tv $, $a>1$&$5\frac 23-4a+\omega_2$&$ q^n$&$1/c^{2(n+1)}$ \\
\hline
\it 35a
&$v'\theta^2 {\cal D}_tv $, $a<1$&$1\frac 23-4a+\omega_2$&$ q^n$&$1/c^{2(n+1)}$ \\
\hline
\it 36
&$v'\theta^2 \Delta v $, $a>1$&$5\frac 23-4a+\omega_2$&$ q^n$&$1/c^{2(n+1)}$ \\
\hline
\it 36a
&$v'\theta^2 \Delta v $, $a<1$&$1\frac 23-4a+\omega_2$&$ q^n$&$1/c^{2(n+1)}$ \\
\hline
\it 37
&$v'\theta^2 (\theta\partial)\theta$, $a<1$&$5\frac 23-4a+2\omega_2$&$ q^n$
&$1/c^{2(n+1)}$ \\
\hline
\it 37a
&$v'\theta^2 (\theta\partial)\theta$, $a>1$&$1\frac 23-4a+2\omega_2$&$ q^n$
&$1/c^{2(n+1)}$ \\
\hline
\it 38
&$\theta^2\partial {\cal D}_tF_p$&$3-4a+\omega_2$&$u',\quad q^n$
&$1/{c^{2(n+1)}}$ \\
\hline
\it 39
&$v'\theta^2\Delta$, $ a>1$&$8\frac 23-4a+\omega_2$&$u,\quad q^n$&$1/c^{2(n+2)}$ \\
\hline
\it 39a
&$v'\theta^2\Delta$, $a<1$&$4\frac 23+\omega_2$&$u,\quad q^n$&$1/c^{2(n+2)}$ \\
\hline
\it 40
&$v'\theta^2\Delta F_1$, $a>1$&$7\frac 23-4a+\omega_2$&$ q^n$&$1/c^{2(n+1)}$ \\
\hline
\it 40a
&$v'\theta^2\Delta F_1$, $a<1$&$3\frac 23+\omega_2$&$ q^n$&$1/c^{2(n+1)}$ \\
\hline
\it 41
&$\theta^2\partial F_2$&$2\frac 23 -4a+\omega_2$&$u',
\quad u$&$1/{c^4}$ \\
\hline
\it 42
&$\partial {\cal D}_t\theta^2$&$3\frac 23-4a+\omega_2$&
$u'$&$1/c^2$ \\
\hline
\it 43
&$\partial\theta^2$&$3-4a+\omega_2$&$u',\quad  q$&$1/c^2$ \\
\hline
\it 44
&$\theta'\theta\partial F_2\theta^2$&$5\frac 23-4a+\omega_2$
&$-$&$1/c^2$ \\
\hline
\end{tabular}
 }
\newpage
{
{\bf Table 2}
\ \ \

\ \ \
\begin{tabular}{|c||c||c|c|}
\hline
\multicolumn{4}{|c|}{\sc Comparison of operators of transversal fields}
  \\ \hline
              \hline
\multicolumn{1}{|c|}{\bf N.}&
\multicolumn{1}{|c|}{\bf Essential operators }&
\multicolumn{1}{|c|}{\bf Unessential operators}&
\multicolumn{1}{|c|}{\bf Comments} \\
  \hline
\it 1
&  $NN^0 3,4$&$NN^0 6,10$&$-$  \\             \hline
\it 2
&  $N^0 25$&$NN^0 3,4$&$n=1$  \\             \hline
\it 3
&  $N^0 1$&$NN^0 8$&$n=0$  \\             \hline
\it 4
&  $N^0 12$&$NN^0 2,28$&$n=0$  \\             \hline
\it 5
&  $N^0 16$&$N^0 5$&$-$  \\             \hline
\it 6
&  $N^0 19$&$N^0 23$&$n=0,\quad l=1$  \\             \hline
\it 7
&  $NN^0 12,14$&$NN^0 30$&$a<2/3+1/4\omega_2$  \\             \hline
\it 8
&  $N^0 30$&$NN^0 12,14$&$a>2/3+1/4\omega_2$  \\             \hline
\it 9
&  $NN^0 16,18$&$N^0 42$&$a<2/3+1/4\omega_2$  \\             \hline
\it 10
&  $N^0 42$&$NN^0 16,18$&$a>2/3+1/4\omega_2$  \\             \hline
\it 11
&  $N^0 19$&$N^0 41$&$a<2/3+1/4\omega_2$  \\             \hline
\it 12
&  $N^0 41$&$N^0 19$&$a>2/3+1/4\omega_2$  \\             \hline
\it 13
&  $N^0 21$&$N^0 32$&$l=1,\quad a<2/3+1/4\omega_2$  \\             \hline
\it 14
&  $N^0 32$&$N^0 21$&$l=1,\quad a>2/3+1/4\omega_2$  \\             \hline
\it 15
&  $N^0 24$&$N^0 31$&$n=1,\quad a<2/3+1/4\omega_2$  \\             \hline
\it 16
&  $N^0 31$&$N^0 24$&$n=1,\quad a>2/3+1/4\omega_2$  \\             \hline
\it 17
&  $N^0 31$&$N^0 34$&$-$  \\             \hline
\it 18
&  $N^0 1$&$N^0 39,39a$&$n=0$  \\             \hline
\it 19
&  $N^0 27$&$N^0 43$&$ a<2/3+1/4\omega_2$  \\             \hline
\it 20
&  $N^0 43$&$N^0 27$&$ a>2/3+1/4\omega_2$  \\             \hline
\it 21
&  $N^0 25$&$N^0 29$&$-$  \\             \hline
\it 22
&  $N^0 35a-36a$&$N^0 40$&$-$  \\             \hline
\it 23
&  $N^0 35a-36a$&$N^0 37(-a)$&$-$  \\             \hline
\end{tabular}
}
\newpage

{\small
\unitlength=1mm
\special{em:linewidth 0.4pt}
\linethickness{0.4pt}
\begin{picture}(148.00,28.00)
\emline{34.00}{28.00}{1}{20.00}{14.00}{2}
\emline{20.00}{14.00}{3}{34.00}{0.00}{4}
\emline{29.00}{5.00}{5}{29.00}{23.00}{6}
\emline{23.00}{9.00}{7}{25.00}{12.00}{8}
\emline{68.00}{28.00}{9}{54.00}{14.00}{10}
\emline{54.00}{14.00}{11}{68.00}{0.00}{12}
\emline{63.00}{5.00}{13}{63.00}{23.00}{14}
\emline{57.00}{9.00}{15}{59.00}{12.00}{16}
\emline{103.00}{28.00}{17}{89.00}{14.00}{18}
\emline{89.00}{14.00}{19}{103.00}{0.00}{20}
\emline{98.00}{5.00}{21}{98.00}{23.00}{22}
\emline{92.00}{9.00}{23}{94.00}{12.00}{24}
\emline{139.00}{28.00}{25}{125.00}{14.00}{26}
\emline{125.00}{14.00}{27}{139.00}{0.00}{28}
\emline{134.00}{5.00}{29}{134.00}{23.00}{30}
\emline{128.00}{9.00}{31}{130.00}{12.00}{32}
\emline{127.00}{19.00}{33}{127.00}{16.00}{34}
\emline{127.00}{16.00}{35}{129.00}{16.00}{36}
\emline{91.00}{16.00}{37}{94.00}{16.00}{38}
\emline{91.00}{16.00}{39}{91.00}{19.00}{40}
\emline{56.00}{19.00}{41}{56.00}{16.00}{42}
\emline{56.00}{16.00}{43}{59.00}{16.00}{44}
\emline{22.00}{19.00}{45}{22.00}{16.00}{46}
\emline{22.00}{16.00}{47}{24.00}{16.00}{48}
\emline{28.00}{17.00}{49}{29.00}{20.00}{50}
\emline{29.00}{20.00}{51}{30.00}{17.00}{52}
\emline{31.00}{1.00}{53}{30.00}{4.00}{54}
\emline{30.00}{4.00}{55}{33.00}{3.00}{56}
\emline{65.00}{28.00}{57}{65.00}{25.00}{58}
\emline{65.00}{25.00}{59}{67.00}{25.00}{60}
\emline{58.00}{7.00}{61}{61.00}{7.00}{62}
\emline{61.00}{7.00}{63}{61.00}{9.00}{64}
\emline{99.00}{17.00}{65}{98.00}{20.00}{66}
\emline{98.00}{20.00}{67}{97.00}{16.00}{68}
\emline{100.00}{1.00}{69}{99.00}{4.00}{70}
\emline{99.00}{4.00}{71}{102.00}{3.00}{72}
\emline{137.00}{28.00}{73}{137.00}{26.00}{74}
\emline{137.00}{26.00}{75}{140.00}{26.00}{76}
\emline{132.00}{9.00}{77}{133.00}{6.00}{78}
\emline{133.00}{6.00}{79}{131.00}{6.00}{80}
\put(124.00,13.00){\rule{1.00\unitlength}{2.00\unitlength}}
\put(88.00,13.00){\rule{1.00\unitlength}{2.00\unitlength}}
\put(53.00,13.00){\rule{1.00\unitlength}{2.00\unitlength}}
\put(19.00,13.00){\rule{1.00\unitlength}{2.00\unitlength}}
\put(20.00,20.00){\makebox(0,0)[cc]{$\theta'$}}
\put(25.00,24.00){\makebox(0,0)[cc]{$\theta$}}
\put(35.00,26.00){\makebox(0,0)[cc]{$\theta$}}
\put(32.00,19.00){\makebox(0,0)[cc]{$v'$}}
\put(32.00,9.00){\makebox(0,0)[cc]{$v$}}
\put(20.00,8.00){\makebox(0,0)[cc]{$\theta$}}
\put(25.00,4.00){\makebox(0,0)[cc]{$\theta$}}
\put(35.00,3.00){\makebox(0,0)[cc]{$\theta'$}}
\put(54.00,20.00){\makebox(0,0)[cc]{$\theta'$}}
\put(59.00,25.00){\makebox(0,0)[cc]{$\theta$}}
\put(54.00,9.00){\makebox(0,0)[cc]{$\theta$}}
\put(58.00,5.00){\makebox(0,0)[cc]{$\theta'$}}
\put(65.00,20.00){\makebox(0,0)[cc]{$v$}}
\put(65.00,8.00){\makebox(0,0)[cc]{$v$}}
\put(68.00,3.00){\makebox(0,0)[cc]{$\theta$}}
\put(70.00,26.00){\makebox(0,0)[cc]{$\theta'$}}
\put(89.00,20.00){\makebox(0,0)[cc]{$v'$}}
\put(94.00,24.00){\makebox(0,0)[cc]{$v$}}
\put(105.00,26.00){\makebox(0,0)[cc]{$\theta$}}
\put(101.00,20.00){\makebox(0,0)[cc]{$\theta'$}}
\put(101.00,9.00){\makebox(0,0)[cc]{$\theta$}}
\put(90.00,8.00){\makebox(0,0)[cc]{$v$}}
\put(94.00,5.00){\makebox(0,0)[cc]{$v$}}
\put(103.00,2.00){\makebox(0,0)[cc]{$\theta'$}}
\put(124.00,20.00){\makebox(0,0)[cc]{$v'$}}
\put(130.00,24.00){\makebox(0,0)[cc]{$v$}}
\put(141.00,24.00){\makebox(0,0)[cc]{$\theta'$}}
\put(137.00,20.00){\makebox(0,0)[cc]{$\theta$}}
\put(137.00,11.00){\makebox(0,0)[cc]{$\theta$}}
\put(141.00,3.00){\makebox(0,0)[cc]{$\theta$}}
\put(125.00,9.00){\makebox(0,0)[cc]{$v$}}
\put(129.00,5.00){\makebox(0,0)[cc]{$v'$}}
\put(145.00,13.00){\makebox(0,0)[cc]{,}}
\end{picture}
}
\vspace{10cm}

 \begin{center}
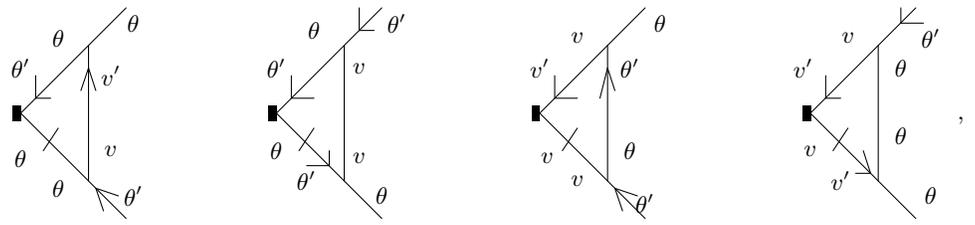
{\bf figure 1.} \hspace{0.5cm} The diagrams  for renormalization of
family of operators of \\
$ \partial\phi'\phi $ -type.
 \end{center}

\newpage

{\small
\unitlength=1mm
\special{em:linewidth 0.4pt}
\linethickness{0.4pt}
\begin{picture}(150.00,42.00)
\put(18.00,23.00){\rule{1.00\unitlength}{2.00\unitlength}}
\emline{6.00}{24.00}{1}{18.00}{24.00}{2}
\emline{19.00}{25.00}{3}{36.00}{42.00}{4}
\emline{19.00}{22.00}{5}{36.00}{5.00}{6}
\emline{32.00}{9.00}{7}{32.00}{38.00}{8}
\emline{12.00}{26.00}{9}{14.00}{24.00}{10}
\emline{14.00}{24.00}{11}{12.00}{23.00}{12}
\put(56.00,23.00){\rule{1.00\unitlength}{2.00\unitlength}}
\emline{44.00}{24.00}{13}{56.00}{24.00}{14}
\emline{57.00}{25.00}{15}{74.00}{42.00}{16}
\emline{57.00}{22.00}{17}{74.00}{5.00}{18}
\emline{70.00}{9.00}{19}{70.00}{38.00}{20}
\emline{50.00}{26.00}{21}{52.00}{24.00}{22}
\emline{52.00}{24.00}{23}{50.00}{23.00}{24}
\put(91.00,23.00){\rule{1.00\unitlength}{2.00\unitlength}}
\emline{79.00}{24.00}{25}{91.00}{24.00}{26}
\emline{92.00}{25.00}{27}{109.00}{42.00}{28}
\emline{92.00}{22.00}{29}{109.00}{5.00}{30}
\emline{105.00}{9.00}{31}{105.00}{38.00}{32}
\emline{85.00}{26.00}{33}{87.00}{24.00}{34}
\emline{87.00}{24.00}{35}{85.00}{23.00}{36}
\put(129.00,23.00){\rule{1.00\unitlength}{2.00\unitlength}}
\emline{117.00}{24.00}{37}{129.00}{24.00}{38}
\emline{130.00}{25.00}{39}{147.00}{42.00}{40}
\emline{130.00}{22.00}{41}{147.00}{5.00}{42}
\emline{143.00}{9.00}{43}{143.00}{38.00}{44}
\emline{123.00}{26.00}{45}{125.00}{24.00}{46}
\emline{125.00}{24.00}{47}{123.00}{23.00}{48}
\emline{23.00}{31.00}{49}{26.00}{32.00}{50}
\emline{26.00}{32.00}{51}{26.00}{29.00}{52}
\emline{26.00}{17.00}{53}{28.00}{13.00}{54}
\emline{28.00}{13.00}{55}{25.00}{13.00}{56}
\emline{64.00}{13.00}{57}{67.00}{12.00}{58}
\emline{67.00}{12.00}{59}{66.00}{15.00}{60}
\emline{69.00}{26.00}{61}{70.00}{30.00}{62}
\emline{70.00}{30.00}{63}{71.00}{26.00}{64}
\emline{96.00}{31.00}{65}{100.00}{33.00}{66}
\emline{100.00}{33.00}{67}{98.00}{29.00}{68}
\emline{98.00}{18.00}{69}{101.00}{13.00}{70}
\emline{101.00}{13.00}{71}{97.00}{15.00}{72}
\emline{133.00}{17.00}{73}{137.00}{15.00}{74}
\emline{137.00}{15.00}{75}{135.00}{19.00}{76}
\emline{142.00}{24.00}{77}{143.00}{29.00}{78}
\emline{143.00}{29.00}{79}{144.00}{24.00}{80}
\put(8.00,28.00){\makebox(0,0)[cc]{$v'_i$}}
\put(47.00,28.00){\makebox(0,0)[cc]{$v'_i$}}
\put(81.00,29.00){\makebox(0,0)[cc]{$v'_i$}}
\put(120.00,30.00){\makebox(0,0)[cc]{$v'_i$}}
\put(19.00,30.00){\makebox(0,0)[cc]{$\theta$}}
\put(26.00,38.00){\makebox(0,0)[cc]{$\theta'$}}
\put(39.00,41.00){\makebox(0,0)[cc]{$\theta$}}
\put(35.00,33.00){\makebox(0,0)[cc]{$v$}}
\put(35.00,15.00){\makebox(0,0)[cc]{$v$}}
\put(39.00,6.00){\makebox(0,0)[cc]{$\theta$}}
\put(19.00,14.00){\makebox(0,0)[cc]{$\theta$}}
\put(26.00,9.00){\makebox(0,0)[cc]{$\theta'$}}
\put(58.00,31.00){\makebox(0,0)[cc]{$\theta$}}
\put(65.00,39.00){\makebox(0,0)[cc]{$\theta$}}
\put(58.00,16.00){\makebox(0,0)[cc]{$\theta'$}}
\put(65.00,10.00){\makebox(0,0)[cc]{$\theta'$}}
\put(77.00,41.00){\makebox(0,0)[cc]{$\theta$}}
\put(77.00,7.00){\makebox(0,0)[cc]{$\theta$}}
\put(73.00,32.00){\makebox(0,0)[cc]{$v'$}}
\put(73.00,18.00){\makebox(0,0)[cc]{$v$}}
\put(93.00,32.00){\makebox(0,0)[cc]{$v$}}
\put(100.00,38.00){\makebox(0,0)[cc]{$v'$}}
\put(94.00,16.00){\makebox(0,0)[cc]{$v$}}
\put(100.00,11.00){\makebox(0,0)[cc]{$v'$}}
\put(112.00,41.00){\makebox(0,0)[cc]{$\theta$}}
\put(112.00,6.00){\makebox(0,0)[cc]{$\theta$}}
\put(107.00,30.00){\makebox(0,0)[cc]{$\theta$}}
\put(108.00,15.00){\makebox(0,0)[cc]{$\theta$}}
\put(132.00,31.00){\makebox(0,0)[cc]{$v$}}
\put(137.00,37.00){\makebox(0,0)[cc]{$v$}}
\put(150.00,39.00){\makebox(0,0)[cc]{$\theta$}}
\put(146.00,32.00){\makebox(0,0)[cc]{$\theta'$}}
\put(146.00,17.00){\makebox(0,0)[cc]{$\theta$}}
\put(132.00,16.00){\makebox(0,0)[cc]{$v$}}
\put(138.00,10.00){\makebox(0,0)[cc]{$v'$}}
\put(150.00,4.00){\makebox(0,0)[cc]{$\theta$}}
\end{picture}
}
\vspace{10cm}
 \begin{center}
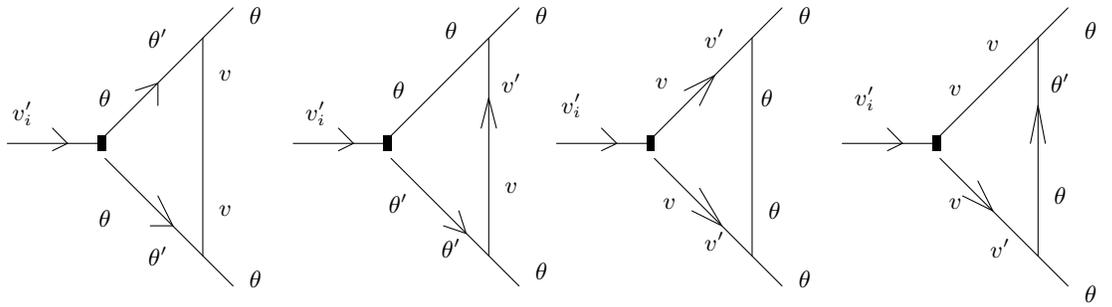
{\bf figure 2.} \hspace{0.5cm} The diagrams  for renormalization of
family of operators of \\
$ \phi'\phi^{2} $ -type.
 \end{center}
\end{document}